\documentstyle[12pt,aas2pp4]{article}

\begin{document}

\title{ON THE ORIGIN OF SOLAR OSCILLATIONS}

\author{Philip R. Goode$^1$ and Louis H. Strous$^2$}
\affil{Big Bear Solar Observatory$^1$, New Jersey Institute of Technology$^2$, Newark, New Jersey, 07102 USA}

\author{Thomas R. Rimmele}
\affil{\altaffilmark{3} National Solar Observatory, Sunspot, New Mexico, 88349 USA}
  
\altaffiltext{3} {Operated by the Association of Universities
for Research in Astronomy, Inc. (AURA) under cooperative agreement
with the National Science Foundation}

\author{Robin T. Stebbins}
\affil{Joint Institute for Laboratory Astrophysics CB\#440, University of Colorado, Boulder, Colorado, 80309-0440 USA}

\begin{abstract}
We have made high resolution observations of the Sun in which we
identify individual sunquakes and see power from these seismic events being
pumped into the resonant modes of vibration of the Sun.   

A typical event lasts about five minutes.  We report the 
physical properties of the events and relate them to theories of the
excitation of solar oscillations.  We also discuss the local seismic 
potential of these events.
\end{abstract}


\section{Introduction}
Earthquakes shake the Earth allowing one to seismically sound our planet's 
interior.  In an analogous way, sunquakes enable a sounding of the solar 
interior.  The two kinds of quakes are also alike in that both are near
surface phenomena.  However, sunquakes are occurring somewhere on the Sun 
all the time, so that energy is being continuously fed to the Sun's resonant
modes implying that one, in principle, could continuously sound the Sun. From 
the global solar seismic data, we have learned a great deal about the Sun's 
interior.  However, the precise origin of individual sunquakes has been 
shrouded in mystery, unlike the origin of earthquakes.

The resonant or normal modes of vibration of the Sun are compressional 
waves.  These so-called p-modes have been long known to have a period 
of about five minutes, and in the solar photosphere they are evanescent 
vertically and traveling horizontally.  It is generally agreed that 
these solar oscillations are excited near the Sun's surface by convection.  
Until recently, it was widely believed that this deceleration of the upgoing 
granules induced a steady drumming that fed the resonant acoustic modes.  
However, \markcite{rimm1995} Rimmele, et al.\ (1995) observed that there are seismic events 
which they associated with the excitation of solar oscillations.  These events 
originate in the dark intergranular lanes.  Furthermore, they observed 
that the seismic events were preceded by a further darkening of an already 
dark lane, and on the temporal leading edge of the seismic event there 
is a still further, and more abrupt darkening.  From this, Rimmele, et al.  
suggested that the excitation of the resonant modes was caused by the 
occasional, catastrophic cooling and collapse of the lanes.  However, they were
unable to show a causal link between the seismic events and the resonant 
modes of vibration of the Sun.

In establishing the causal link here, we show that a seismic event is a 
local power surge that feeds the normal modes.

\section{The Data}

Our observations were made at the Vacuum Tower Telescope of the National Solar 
Observatory in Sunspot, New Mexico.  The quiet Sun dataset discussed here is 
from Sept. 5, 1994.  The data and the reduction of it are described in 
detail in \markcite{rimm1994} Rimmele, et al.\ (1994, 1995), and references therein.  One of the basic 
observational problems here is to distinguish the seismic event power from the 
dominant resonant mode power.  To distinguish the two in our field of 
view (60"x60" patch of quiet Sun near disk center), Rimmele, et al.  measured 
the velocity as a function of altitude in the photosphere for 65 min by 
observing the Doppler shift in the 543.4 nm Fe I absorption line.  The Doppler 
shift as a function of depth in the line corresponds to the velocity as a 
function of altitude in the atmosphere the line spans.  In our analyses here, 
we further explore the data of Rimmele, et al.

We note that earlier spectrographic observations of the Fe I line,
\markcite{steb1987} Stebbins and Goode (1987), but with a faster cadence and one dimensional 
field of view, the basic life cycle of the events was the same ours, see 
\markcite{rest1993} Restaino, et al. (1993).  Also, observations measuring other lines show 
the same basic results for the events, e.g., the well-known Fe I 5576
\AA (g=0) line.

\section{The Seismic Events}

We searched our velocity field for phase changes with altitude and
found they fit one behavior pattern$--$uniformly looking like an
outgoing wave followed by a wave coming back down from above (with a
time lag of about 4 to 5 minutes). The signature of these seismic events 
was detected in the solar photosphere which is not quite isothermal implying 
that any outgoing wave would be followed by a partially reflected wave.  
This combination of phase behaviors eased our effort to distinguish 
power from seismic events from that of normal modes which should show 
only a small phase change with altitude caused by dissipation,
Restaino, et al. (1993)

In the analysis, we superposed slightly more than two thousand separate 
seismic events each of which constitutes an event covering more than
100 voxels (pixels in all frames combined).  This is a convenient and efficient
way to separate significant seismic events from background noise.
The superposed events were pinned in time with $T=0$ being the peak in 
the product of the square of the acoustic  velocity and the vertical 
phase gradient  for each event. After superposing the seismic events, 
each was oriented such that the intergranular lane was along the x-axis, see 
the three panels of Figure 1.  If the events were purely traveling acoustic 
waves, the aforementioned product would be proportional to the 
acoustic or mechanical flux.  Regardless, the product is a 
convenient measure of seismic events.  Our use of standard formulae 
for calculating the mechanical flux of  linear, traveling 
waves predicts energy deposition which may 
be no better than to the order of magnitude for the waves considered
here.  After all, the waves from the sunquakes may not be fully linear, 
and are not waves supported by the atmosphere, rather they are the result of a
temporal forcing of the photosphere.

\begin{figure*}[t]
\vspace{-4cm}
\epsscale{2}
\plotone{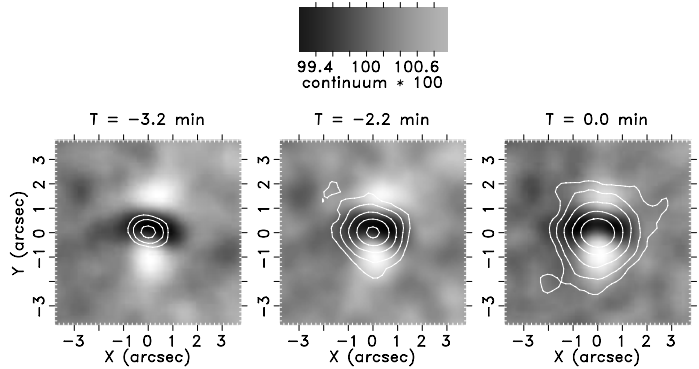}
\vspace{-10cm}
\caption[fig1.eps]{\label{fig1}
The superposed seismic flux, shown in white contours, is 
superposed on its local, averaged  granulation altitude 150 km
above the base of the photosphere.  The contours are 0.25, 0.5, 1, 
2, 4 $\times10^7$ ergs/cm$^2$/s.  The flux is shown for $T=$ $-$3.2, $-$2.2 
and 0 min.  The time steps in the data collection are about 30 s, and
one time step before $T=-$3.2 min, the seismic flux is below 
0.25 $\times10^7$ ergs/cm$^2$/s everywhere in the field of view .
}
\end{figure*}

In Figure 1, $T=0$ is the time of peak ``seismic flux", and the seismic flux 
contours in Figure 1 are normalized to the mechanical flux as defined
in Rimmele, et al.  The fact that a dark lane with a bright granule on 
either side  survives the averaging strongly emphasizes this geometry is a 
common feature of seismic events.  The granular contrast is small
because granular images are smeared by the averaging. It is also
obvious that more than 2" from the center of the field of view, the granular 
structure is completely washed out by the averaging of many hundreds 
of events.  In the figure, the bright contours represent the outgoing seismic 
flux.  Clearly, the seismic events originate in the lanes, for more detail see 
 Rimmele, et al.\ (1995).

Further, immediately after the peak in the seismic flux, the lane begins to 
narrow as though the granules on either side of the lane are being pulled 
together to fill the void left behind.  

From Figure 1, it is also clear that seismic events have a finite duration.  
Over the three minute span shown, increasing seismic energy can be seen 
being fed into the aggregated events.  After $T$=0 in the figure, the 
seismic flux gradually subsides.  The total duration of the expansive 
phase of the event is about five minutes.  The apparent 
finite duration of the events needs to be emphasized for several
reasons.  The fact that the peak in the observed spectrum of global 
solar oscillations corresponds to modes with a period of about 5 minutes may
well be connected to the comparable temporal duration of the seismic
events.  That is, because the events are not impulsive, and, in fact
endure for a time comparable to the period of the oscillations,
resonance may play a role in the excitation of the oscillations.
The conventional wisdom has been, however, that turbulent convection excites
acoustic waves in a broad frequency range, including waves above the
acoustic cut-off.  

Arguments for seismic power above the cut-off are predicated on the assumption 
that the excitation is nearly instantaneous so that the power in the stably 
stratified solar photosphere will be above the acoustic cut-off.  
\markcite{lamb1909} Lamb (1909) first idealized the description of a seismic event by considering
the response of an infinite, stably stratified isothermal atmosphere to a 
thumping from below.  If an impulsive, Lamb-like picture were correct,  
then observers could employ simpler techniques than we did.  Observers
could simply look for the signature of seismic events above the acoustic 
cut-off, instead of below the acoustic cut-off where the power is dominated 
by p-modes.  Such an approach was taken by \markcite{brow1992} Brown, et al. (1992). but
like them, we find no appreciable seismic event power in the quiet Sun 
above the cut-off.  This is not a great surprise since the seismic flux
power spectrum has a fairly narrow peak near a five minute period.  Thus, to 
look for seismic event power, one has to look where the power is$--$this 
means having to distinguish seismic events from resonant 
modes.  As discussed, we do this by identifying individual events by their
characteristic large phase shift with altitude which is a strong
function of time.  

The five minute timescale of the events is consistent with that calculated 
in a linear, one dimensional model of seismic events,\markcite{good1992} Goode, Gough and 
Kosovichev (1992).  In their simulations, they showed that the mean
velocity and phase properties of data like that of Rimmele, et al. are 
described well only if the typical event endures for about five minutes
at its subsurface point of origin immediately beneath the base of the 
photosphere.  They demonstrated, for instance, if the sub-photospheric 
model disturbance were more impulsive, the model signal would be too
impulsive in the photosphere to describe the data.  Updating the simulations 
of Goode, Gough and Kosovichev we find that, in the mean, the calculated
flux (from the work done by the forcing) matches that observed (seismic
flux from the data) best by making the assumption that the  speed 
of the convective downdraft is close to 1 km/sec at the photosphere.

\section{Powering the Oscillations}

In Figure 2, we show the superposition of the power at 150 km above the
base of the photosphere and the instantaneous phase difference 
between that altitude and 180 km higher. The specific model altitudes 
were provided by \markcite{keil1997} Keil (1997, private communication).  Both quantities 
in Figure 2 are shown as a function of time and horizontal distance 
from the event with $T=0$ being the peak in the seismic flux for the 
superposed events.  We remark that what is generally regarded as 
being convective power is subsonic and has been filtered out.  
The $k-\omega$ diagram for our data is shown in Figure 3.  

\begin{figure*}[t]
\vspace{-6.cm}
\epsscale{3}
\centerline{\plotone{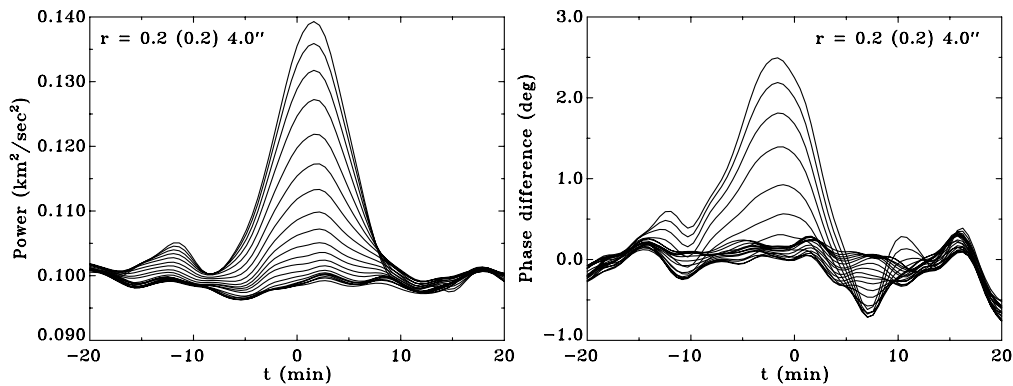}}
\vspace{-19cm}
\caption[fig2.eps]{\label{fig2}
a) The average excess power ($v^2$) in the neighborhood 
of more than two thousand seismic events, superposed as in Figure 1, are 
as shown as a function of distance and time$--$starting at r=0.2" (the curve 
with the highest peak power) in steps of 0.2" out to 4.0".  Successive 
distances show successively decreasing peak power out to about 3.0".
b)  The averaged phase difference between two altitudes in the 
photosphere ( 150 km and 330 km above the base of the photosphere) as 
a function of time and distance in steps of 0.2" as in a).  The formal
standard error on the averaged phase difference is about 0.1$\deg$ at
each spatial and temporal point.  The
largest positive phase change is for r=0.2".  A positive phase change 
corresponds to an upgoing wave. Successive distances show successively
decreasing peak positive phase change out to about 1.4".  A
typical event has a peak phase difference significantly larger than
that of the superposed, and therefore smeared result in the figure.
}
\end{figure*}

\begin{figure*}[t]
\vspace{-2cm}
\epsscale{2.5}
\centerline{\plotone{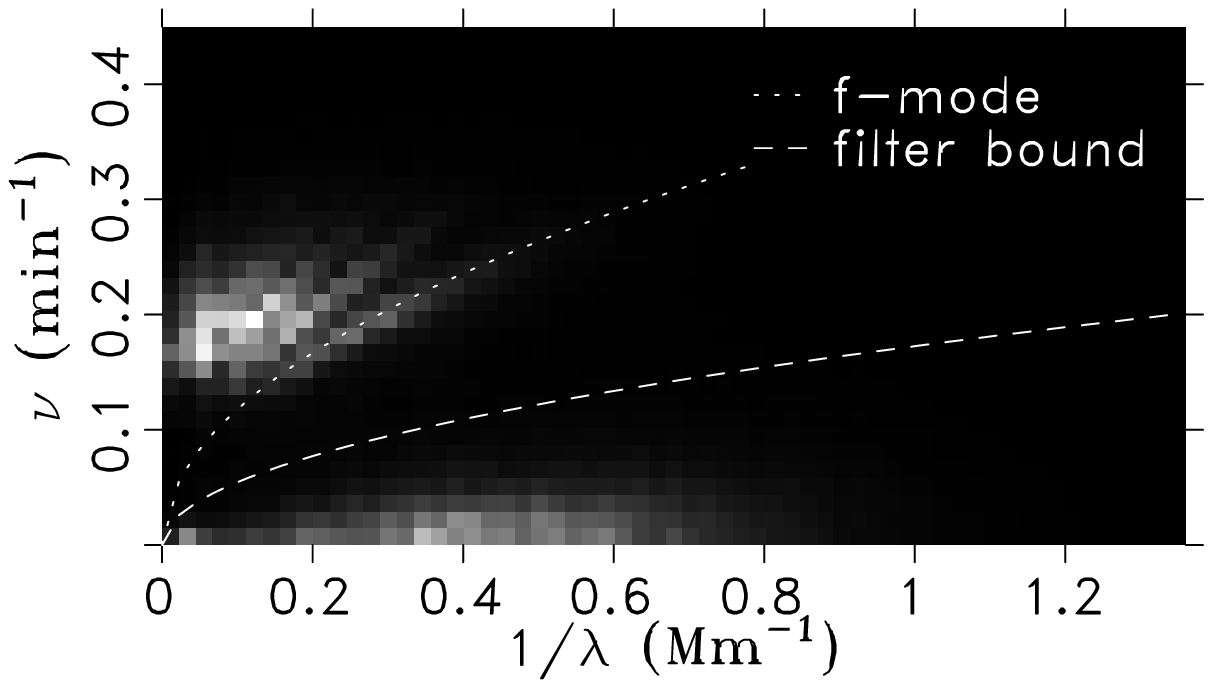}}
\vspace{-13cm}
\caption[fig3.eps]{\label{fig3}
The $k-\omega$ diagram for our data.  Model
predictions for the sonic cut-off and f-mode ridge are indicated by the
bold white curves.  Note that the sonic and subsonic powers are 
well-separated.  Note also, that there is significant 
power in the region of the f-mode ridge .
}
\end{figure*}

The phase signature characteristic of seismic events is apparent clear out 
to about 1.4'' from the events.  Beyond that distance, the phase change
with altitude is essentially zero.  However, there is excess power from 
the events going out almost 3''.  Beyond that distance, no excess power 
is apparent. The tendency of the power is to decrease as the square of the 
distance from the events.  This tendency is what one might anticipate.
The power propagates with an apparent supersonic speed out to about
1.5'', but in all liklihood, that speed reflects the horizontal extent
of the events rather than a true propagation speed
(Imagine a piston emerging from a fluid -- the horizontal speed
measured from the axis to the edge of the piston only reflects the
shape of the top of the piston , Strous, et al. (1997)).  
Beyond 1.5'', and beyond the edge of the events, there is a true
propagation speed which is slightly supersonic.  It is likely that 
most of this power is in f-modes which are {\it asymptotically} (in terms of
horizontal wavelength) surface waves.  We first note that the region of
the f-mode ridge is apparent in our $k-\omega$ diagram (see Figure 3).
The f-modes form the lowest frequency ridge in the p-mode $k-\omega$ diagram.  
Physically, they are somewhat distinct from the rest.  Two particularly 
distinct properties of the f-modes is that they are {\it asymptotically} 
incompressible and may propagate horizontally with a slightly 
supersonic group velocity.

Our contention that power has been fed from seismic events into the f-mode 
part of the spectrum of solar oscillations is greatly strengthened 
because the acoustic power delivered by the events to beyond 1.4'' is: 
1) characterized by no vertical phase change
which means an essentially infinite vertical phase speed, 2) characterized 
by a five minute period and a group velocity roughly appropriate for f-modes, 
and 3) dominated by horizontal wavelengths consistent with those of the 
f-modes.  Thus, power has been fed from seismic events into the Sun's
normal modes.

The events occur just beneath the photosphere, and what we observe is the
photospheric effluevia which is converted into atmospheric p-modes.
However, much of the energy from these events is directed into the Sun.  
The process by which this latter energy is converted into p-modes is 
somewhat different.  The inward directed noise is eventually refracted back to 
the surface where it is partially reflected back into the Sun.  
\markcite{kuma1993} \markcite{kuma1994} 
Kumar (1993, 1994) has shown theoretically that after only a few refractions 
white noise would be converted into resonant modes.  We cannot expect to 
detect the signature of such skips from our dataset, since the typical
distance for a single refraction of, say a five minute period p$_1$ mode 
is much greater in extent on the solar surface than our field of view.
Thus, we don't (and can't) see power being pumped into all modes,
but we see power pumped into a part of the spectrum$--$the f-mode part
for which the skip times are very short.  The seismic events can power
the entire p-mode spectrum: if half the power of the events is fed into 
the Sun's resonant cavity, there is ample power to drive the Sun's entire 
spectrum of oscillations, Rimmele, et al.\ (1995).

Figure 1 also clearly reveals a seismic potential for the events 
low in the photosphere where the convective overshooting occurs.  In 
particular, the events from the lane collapse seem to travel
horizontally about 30\% faster over brighter regions.  Further, 
these seismic data will aid in diagnosing flaws in any forthcoming 
three dimensional model of the events.

\section{Discussion}

In the collapse of an intergranular lane that generates the seismic events, 
one could invoke linear and nonlinear processes.  Linear processes would be
rarefaction waves generated by the collapse and the subsequent downgoing blob
acting like a piston.  Nonlinear ones would be the implosion of the blob on
itself and the infall of material behind the blob (see Rast and Toomre 1993b).

Near the center of a seismic event the disturbance seems quite supersonic, 
which would imply an origin which is, at least, partially nonlinear.
However, this speed probably reflects the finite size of the events, and, thus
provides no evidence of the presence of nonlinear effects.
Nonlinear effects here would take convective power and convert it to
acoustic power.  From the $k-\omega$ diagram, nonlinear effects would
seem to be small since the convective and acoustic powers are well separated.  
Further, the subsonic velocities which in a linear 
theory should correspond to the convective velocities, do in fact 
correspond to the granular velocities we observe.
Thus, the events seem to be nearly linear in origin. 

Following \markcite{gold1977} Goldreich and Keely (1977), the commonly accepted picture of
the excitation of solar oscillations is one in which stochastic driving 
is done by turbulent convection.  This theory has been
further developed by \markcite{gold1988}  Goldreich and Kumar (1988), and has enjoyed
success in explaining the distribution of power within the
p-mode spectrum.  The theory relies, to some extent, on mixing length
formalism in which there is a full symmetry between the role of upgoing and
downgoing flows. However, we clearly see from our seismic events there is
no such symmetry.  To the contrary, strong events occur exclusively in
the dark intergranular lanes.  Thus, we believe that it would be
valuable to account for this asymmetry in a future theoretical 
effort to quantitatively explain the p-mode spectrum.  

Our observations were motivated by the pioneering, large-scale
simulations of convection by \markcite{nord1985} Nordlund (1985) in which he predicted
narrow, supersonic downdrafting plumes.  The simulations of \markcite{rast1993}  Rast and Toomre (1993a,b) predict a role for the plumes in the excitation 
of solar oscillations.  We don't know if the predicted plumes are
associated with the seismic events we observe, but we are working with
Nordlund and co-workers to resolve this.

Seismic events would seem to have the seismic potential to probe the
neighboring granular structure and small scale magnetic fields.  We 
have made observations in regions of weak magnetic structure to determine 
magnetoseismic potential of the seismic events.

\acknowledgments
This work was supported by AFOSR-92-0094 and NAG5-4919.  We gratefully
acknowledge many useful conversations with and suggestions from
W.A. Dziembowski.  We thank Haimin Wang and Scot Kleinman for careful 
readings of the manuscript.

\newpage

\end{document}